# Graph Theory Applications in Network Security

Jonathan Webb1, Fernando Docemmilli2, and Mikhail Bonin3

Theory Lab - Central Queensland University

Wayville SA 5034

E-mail addresses: (1) jonwebb@cqu.edu.au (2) ferdocem@cqu.edu.au (3) mbonin@cqu.edu.au

**Abstract**

Graph theory has become a very critical component in many applications in the computing field including networking and security. Unfortunately, it is also amongst the most complex topics to understand and apply.

In this paper, we review some of the key applications of graph theory in network security. We first cover some algorithmic aspects, then present network coding and its relation to routing.

## 1 INTRODUCTION

The rapid growth in Global mobile communication networks demands new solutions for existing problems. Such problems include reduced bandwidth in mobile devices and the constant change in their associated network topologies. This creates a need for network algorithms with:
1. least possible communication traffic
2. High speed execution.

The two challenges can be overcome by application of graph theory in developing local algorithms (Algorithms that require low rounds of communication). In this paper we explore applications of graph theory in cellular networks with an emphasis on the 'four-color' theorem and network coding and their relevant applications in wireless mobile networks.

## 2 RELATED WORK

Chung and Lu [1] studied the graph theory and it is relation to many practical implementations including security extensively. For example, they investigated power-law models and its relationship with network topologies. They also proved upper and lower bounds to many key computational problems. Ahmat's work [6] focused more on optimization issues related to graph theory and its security applications. He presented some key graph theory concepts used to represent different types of networks. Then he described how networks are modeled to investigate problems related to network protocols. Finally, he presented some of the tools used to generate graph for representing practical networks. This work is considered among the most comprehensive in addressing graph optimization complexity issues in networking and security. More recently, Shirinivas et al. [12] presented an overview of the applications of graph theory in heterogeneous fields to some extent but mainly focuses on the computer science applications that uses graph theoretical concepts.

Network topology discovery has also attracted significant amount of graph theory related research work from academia and industry. In his article [5], K. Ahmat discussed the past and current mechanisms for discovering the Layer-2 network topology from both theoretical and practical prospective. In addition to discovery techniques, he provided some detailed explanations to some of the well known open issues related to Ethernet topology discovery and their use cases. For example, one might want to model the Internet in order to reproduce its behavior in a laboratory.

Breitbart et al. described a pioneer work on Ethernet topology discovery which is critical in network security [2]. They presented novel algorithms for discovering physical topology in heterogeneous (i.e., multi-vendor) IP networks. Their algorithms rely on standard SNMP MIB information that is widely supported by modern IP network elements and require no modifications to the operating system software running on elements or hosts. They also implemented their algorithms in the context of a topology discovery tool that has been tested on Lucent's own research network. The algorithms designed in this paper work only when the MIB information are complete.

Gobjuka and Breitbart [3] addressed the same problem when the information from MIBs are incomplete. Namely, they investigated the problem of finding the layer-2 network topology of large, heterogeneous multisubnet Ethernet networks that may include uncooperative network nodes. They proved that finding a layer-2 network topology for a given incomplete input is an NP-hard problem, even for single subnet networks, and that deciding whether a given input defines a unique network topology is a co-NP-hard problem. They designed several heuristic algorithms to find network topology, evaluate their complexity and provide criteria for instances in which the input guarantees a unique network topology. They also have implemented one of their algorithms and conducted extensive experiments on Kent State University Computer Science network.

There are several researchers who explored graph theory and its practical aspects to Mobile Ad Hoc Networks MANET) too. Saleh Ali K. Al Omari and Putra Sumari [4] presented an exhaustive survey about the Mobile Ad Hoc Network (MANET) and They made a comparison between the different papers, most of its conclusions pointed to a phenomenon, not a routing protocol can adapt to all environments, whether it is Table-Driven, On-Demand or a mixture of two kinds, are limited by the network characteristics. Oliveira et al. [13] proposed a solution for securing heterogeneous hierarchical WSNs with an arbitrary number of levels. Our solution relies exclusively on symmetric key schemes, is highly distributed, and takes into account node interaction patterns that are specific to clustered WSNs.

From privacy prospective, S. Sumathy and B.Upendra Kumar [7] proposed a key exchange and encryption mechanism that aims to use the MAC address as an additional parameter as the message specific key [to encrypt] and forward data among the nodes. In the model they proposed, the nodes are organized in spanning tree fashion, as they avoid forming cycles and exchange of key occurs only with authenticated neighbors in ad hoc networks, where nodes join or leave the network dynamically.

Donnet and Friedman [10] discussed past and current mechanisms for discovering the internet topology at various levels: the IP interface, the router, the AS, and the PoP level. In addition to discovery techniques, they provided insights into some of the wellknown properties of the internet topology. Maarten van Steen [8] focused on the commonly-used measures, namely, those concerning vertex connectivity, small-world property, correlations in connectivity pattern, and centrality. Unless otherwise specified, the complex network measures are of an undirected unweighted linguistic network N = (V, E) with n vertices and m edges as the model of a particular language sub-system.

Breitbart *et al.* [11] described a method for minimizing network monitoring overhead based on Shortest Path Tree (SPT) protocol. They describe two different variations of the problem: the A-Problem and the E-Problem, and show that there is a significant difference between them. They also proved that finding optimal solutions is NP-hard for both variations, and propose a theoretically best possible heuristic for the A-Problem and three different heuristics for the E-Problem.

Patrick P. C. Lee *el al.* [14] proposed a distributed secure multipath solution to route data across multiple paths so that intruders require much more resources to mount successful attacks. They include a distributed routing decisions, bandwidth-constraint adaptation, and lexicographic protection, and proved their convergence to the respective optimal solutions. In his book [15], Remco van der Hofstad studied random graphs as models for real-world networks. He concluded that, these networks turn out to have rather different properties than classical random graph models, for example in the number of connections the elements in the network make. As a result, a wealth of new models was invented so as to capture these properties.

## 3 GRAPH THEORY MODELS IN SECURITY
### 3.1 Graph theory

A graph is a simple geometric structure made up of vertices and lines. The lines may be directed arcs or undirected edges, each linking a pair of vertices. Amongst other fields, graph theory as applied to mapping has proved to be useful in Planning Wireless communication networks.

### 3.2 The Four-Color Graph Theorem

The famous four-color theorem states that for any map, such as that of the contiguous (touching) provinces of France below, one needs only up to four colors to color them such that no two adjacent provinces of a common boundary have the same color. With the aid of computers, mathematicians have been able to prove that this applies for all maps irrespective of the boarder or surface shape

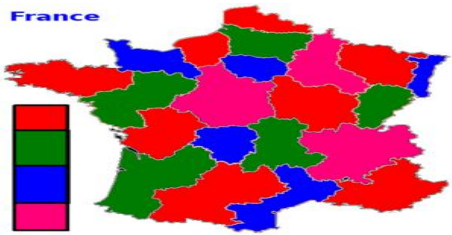

*Fig 1 Provinces of France*

## **Applying of the four color theorem in wireless a cell tower placement plan.**

Consider the cell tower placement map shown above, where each cell tower broadcast channel is likened to a color, and channel–colors are limited to four, the task of finding where to economically position broadcast towers for maximum coverage is equitable to the four-color map problem.

The two challenges are:
1. Elimination of the no-coverage spots ( marked red in the diagram above)
2. Allocation of a different channel in the spots where channel overlap occurs (marked in blue). In analogy, *colors* must be different, so that cell phone signals are *handed off* to a different channel.

Each cell region therefore uses one control tower with a specific channel and the region or control tower adjacent to it will use another tower and another channel. It is not hard to see how by using 4 channels, a node coloring algorithm can be used to efficiently plan towers and channels in a mobile network, a very popular method in use by mobile service providers today

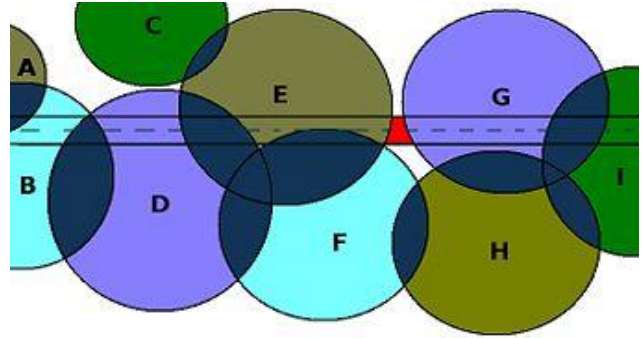

*Fig2. Cellular Mobile tower placement map coloring*

## **Node Coloring Theorem**

As can be seen in the map below, borders wander making it a difficult problem to analyze a map. Instead of using a sophisticated map with many wandering boundaries, it becomes a simpler problem if we use node coloring. If two nodes are connected by a line, then they can't be the same color. Wireless Service providers employ node coloring to make an extremely complex network map much more manageable.

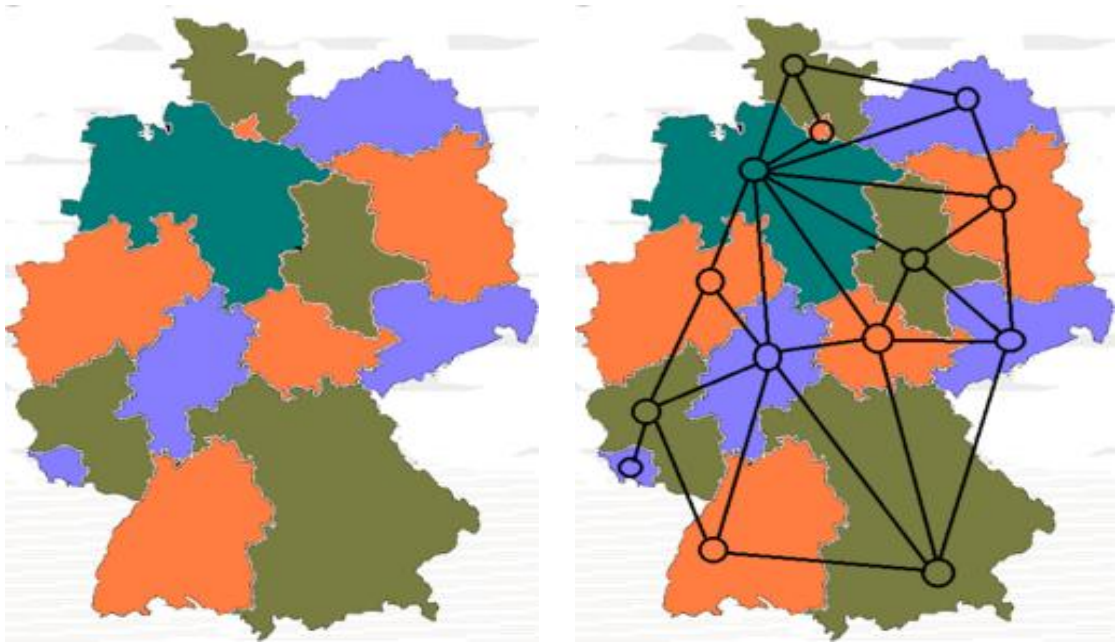

*Fig3(a) A Map with complex wandering boundaries*

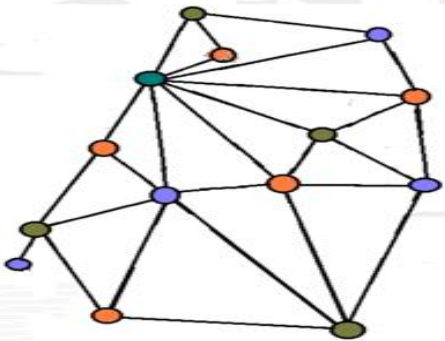

*Fig3(b). The simplified network version of the map derived by node coloring*

## 4   NETWORK CODING

Network coding is another technique where graph theory finds application in mobile communication networks. In a traditional network, nodes can only replicate or forward incoming packets. Using network coding, however, nodes are able to algebraically combine received packets to create new packets

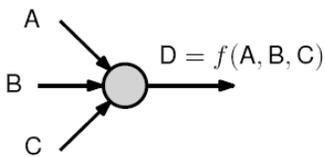

*Fig4. Node replication of forward incoming packets*

Network coding opens up new possibilities in the fields of networking. Such would include:

Wireless multi-hop networks:

- Wireless mesh networks
- Wireless sensor networks
- Mobile ad-hoc networks
- Cellular relay networks

Peer-to-peer file distribution

- Peer-to-peer streaming
- Distributed storage

### Application of network coding in a content distribution scenario

For this application, the following assumptions are made:

1. The network is a multicast system where all destinations wish to receive similar information from the source.
2. That all Links have a unit capacity of a single packet per time slot
3. That the links be directed such that traffic can only flow in one direction.

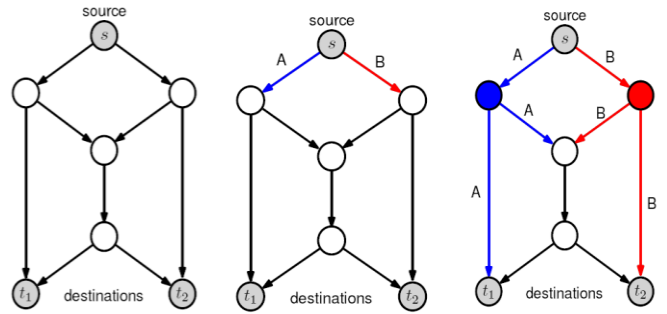

*Fig5   Content distribution scenario*

### 1$^{st}$ time slot

Following the first time slot:

Destination t1 will have received information traffic A whereas Destination t2 will have received both traffic A and traffic B as shown below.

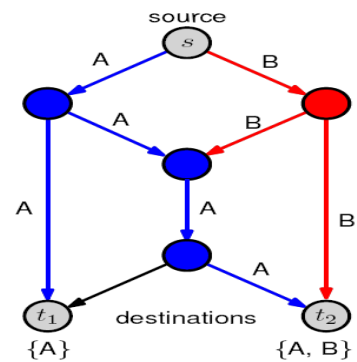

*Fig6 First time slot*

### 2$^{nd}$ time slot

In the second time slot:
Both Destination t1 and t2 will have received traffic A and traffic B and C as shown below

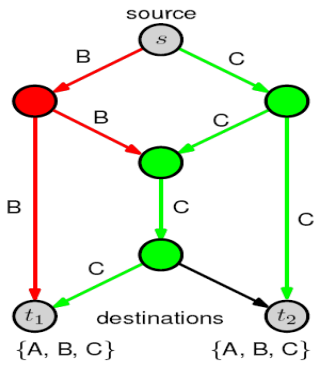

*Fig 7 Second time slot*

Finally, when Destination t1 receives A and A(EX-OR)B, it will be able to compute B by B=A (EX-OR){ A(EX-OR)B}

Likewise, when Destination t2 receives B and A (EX-OR) B, it will be able to compute A by:

A= {A (EX-OR) B} (EX-OR) B as shown below

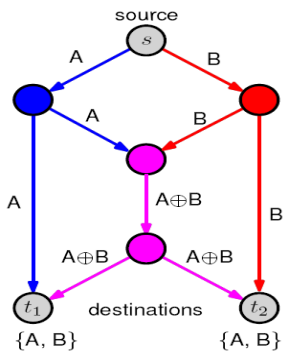

*Fig 8 EX-OR Computation*

## Network Coding application in Opportunistic Routing

Opportunistic routing is a technique that makes use of multiple paths in a network to obtain diversity. Network coding can be applied in such a case (fig 9) to coordinate transmissions in order to avoid duplicate packets

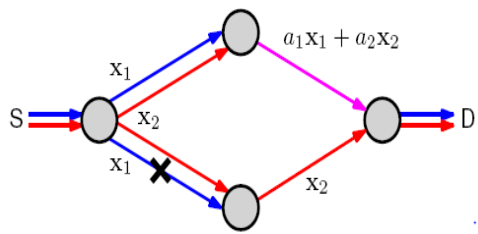

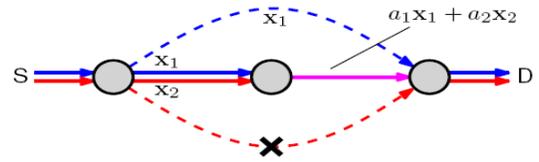

*Fig 10* shows how network coding simply helps us determine how many packets should be sent by each node

## Network coding application in a Physical Layer Network

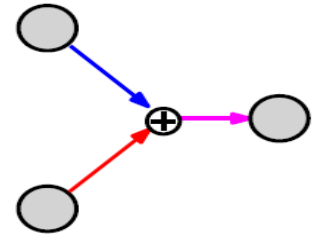

*Fig 11*

Network coding can be used with a physical layer network to enable networks benefit from interference rather than avoid it. Assuming the wireless channel performs network coding over the air. Fig 12 below shows how traditional network coding would perform, while Fig 13 shows physical layer network coding.

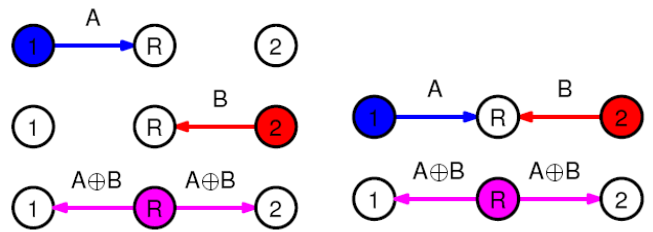

*Fig 12 Traditional N     Fig 13 Fig 12 Physical layer NC*

Here, the challenge is inferring w1 (EX-OR) w2 from the observed signal

$$y = h_1 x_1 + h_2 x_2 + z$$

Special case: $h_1 = h_2 = 1$

Physical-layer Network coding Scheme with BPSK Modulation

This one requires phase synchronization and power control.

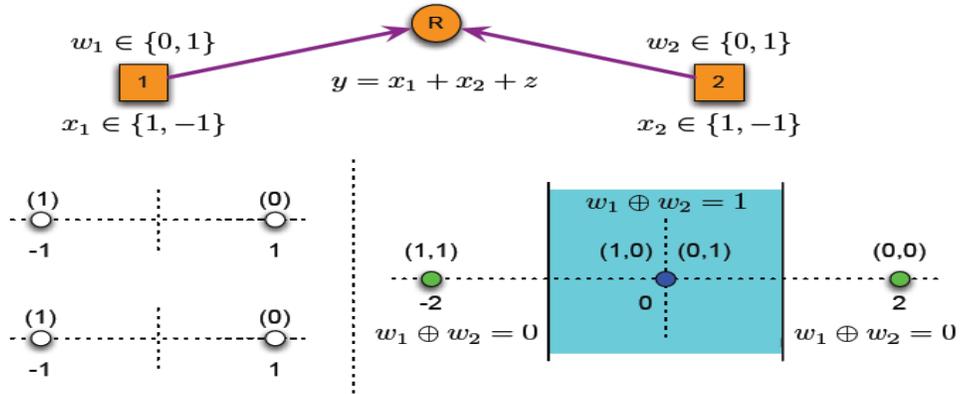

*Fig 14 Physical layer network coding with BPSK Modulation*

Physical-layer Network coding Scheme with QPSK Modulation

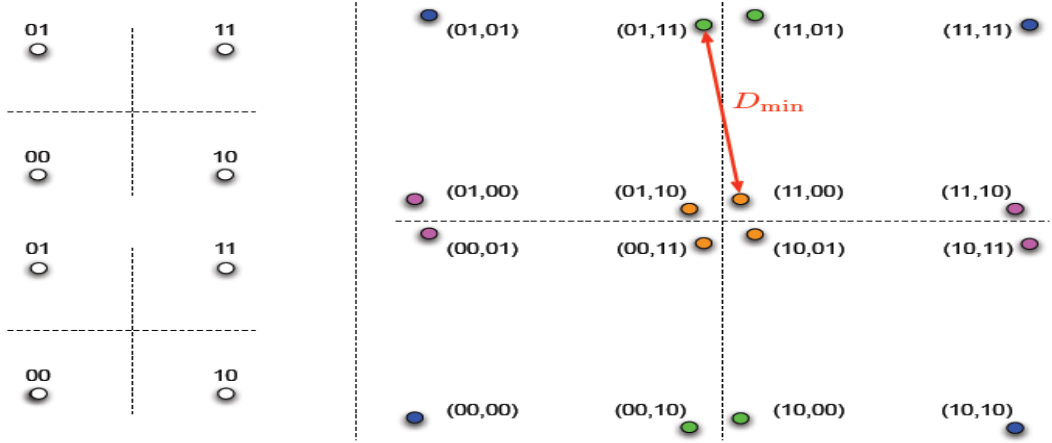

*Fig 16 (a) Physical layer network coding with QPSK Modulation*

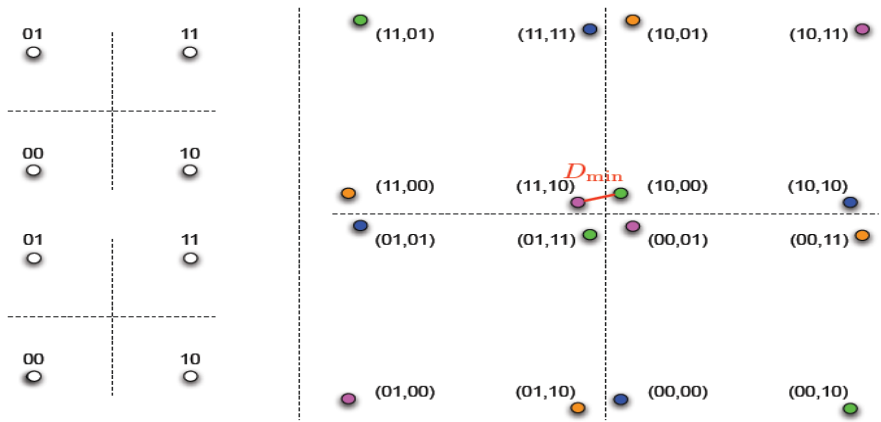

*Fig 16 (b) Physical layer network coding with QPSK Modulation*

# 5   Conclusion

From the examples discussed it was shown that indeed graph theory, as far as the four-color theorem and network coding are concerned, can help provide significant throughput benefits for:

- wireless multi-hop networks
- content distribution scenarios

Other benefits are:

- Time, resource and energy savings

Simplified operation.

## References

[1] F.R.K. Chung and L. Lu. Complex Graphs and Networks, volume 107 of CBMS Regional Conference Series in Mathematics. American Mathematical Society, 2006.

[2] Y. Breitbart, M. Garofalakis, C. Martin, R. Rastogi, S. Seshadri and A. Silberschatz. " Topology Discovery in Heterogeneous IP Networks" In Proceedings of IEEE INFOCOM, 2000.

[3] H. Gobjuka and Y. Breitbart. Ethernet topology discovery for networks with incomplete information. In IEEE/ACM Transactions in Networking, pages 18:1220–1233, 2010.

[4] Saleh Ali K. Al Omari and Putra Sumari, " An overview of Mobile Ad Hoc Networks for the existing protocols and applications", International Journal on applications of graph theory in wireless ad hoc networks and sensor networks, vol. 2, no. 1, March 2010.

[5] K. Ahmat, Ethernet Topology Discovery: A Survey," CoRR, vol. abs/0907.3095, 2009.

[6] K. Ahmat, Graph Theory and Optimization Problems for Very Large Networks. City University of New York, United States, 2009

[7] S. Sumathy et. al. ,"Secure key exchange and encryption mechanism for group communication in wireless ad hoc networks ", Journal on Applications of graph theory in wireless ad hoc networks and sensor networks, Vol 2,No 1,March 2010.

[8] van Steen M. Graph Theory and Complex Networks: An Introduction. Maarten van Steen; 2010.

[9] F.R.K. Chung and L. Lu. Complex Graphs and Networks, volume 107 of CBMS Regional Conference Series in Mathematics. American Mathematical Society, 2006.

[10] B. Donnet, T. Friedman, Internet topology discovery: A survey, Communications Surveys & Tutorials, IEEE 9 (2007) 56–69.

[11] Y. Breitbart, F. Dragan, and H. Gobjuka, "Effective network monitoring," in International Conference on Computer Communications and Networks (ICCCN), 2004.

[12] S. G. Shirinivas, S. Vetrivel, and N. M. Elango, "Applications of graph theory in computer science—an overview," International Journal of Engineering Science and Technology, vol. 2, no. 9, pp. 4610–4621, 2010.

[13] Oliveira, L.B., Wong, H.C., Dahab, R., Loureiro, A.A.F.: On the design of secure protocols for hierarchical sensor networks. International Journal of Security and Networks (IJSN) 2(3/4) (2007) 216–227 Special Issue on ryptography in Networks.

[14] P. C. Lee, V. Misra, and D. Rubenstein. Distributed algorithms for secure multipath routing in attack-resistant networks. IEEE/ACM Transactions on Networking, 15(6):1490–1501, Dec. 2007.

[15] R. van der Hofstad, Random Graphs and Complex Networks, 2011, (manuscript available at: www.win.tue.nl/rhofstad/NotesRGCN2011.pdf).